\documentclass{sig-alternate-05-2015}
\usepackage[utf8]{inputenc}
\usepackage{url}
\usepackage{dirtree}
\usepackage{csquotes}

\usepackage[font=small,labelfont=bf]{caption}

\usepackage{float}
\floatstyle{ruled}
\newfloat{listing}{h}{lop}
\floatname{listing}{Listing}

\usepackage{listings}
\usepackage{etoolbox}
\makeatletter
\patchcmd{\maketitle}{\@copyrightspace}{}{}{}
\makeatother

\usepackage{hyperref}
\hypersetup{
  pdfinfo={
    Author={Pascal Mainini, Annett Laube},
    Title={Access Control in Linked Data Using WebID},
    Producer={},
    Creator={}
  },
  pdfborder={0 0 0},
  colorlinks,
  citecolor=black,
  filecolor=black,
  linkcolor=black,
  urlcolor=black
}

\usepackage{xcolor}

\definecolor{light-gray}{gray}{0.95}

\begin{document}

\setcopyright{rightsretained}

\title{Access Control in Linked Data Using WebID}
\subtitle{A Practical Approach Validated in a Lifelong Learning Use Case}

\numberofauthors{2}
\author{
\alignauthor
Pascal Mainini\\
       \affaddr{Bern University of Applied Sciences (BFH)}\\
			 \affaddr{Institute for ICT Based Management}\\
       \affaddr{Höheweg 80, CH-2502 Biel/Bienne}\\
       \email{pascal.mainini@bfh.ch}
\alignauthor
Prof. Dr. Annett Laube-Rosenpflanzer\\
       \affaddr{Bern University of Applied Sciences (BFH)}\\
			 \affaddr{Institute for ICT Based Management}\\
       \affaddr{Höheweg 80, CH-2502 Biel/Bienne}\\
       \email{annett.laube@bfh.ch}
}
\date{20 April 2016}

\maketitle
\begin{abstract}
Linked Data technologies become increasingly important in many domains. Key factors for their breakthrough are \emph{security and trust},
especially when sensible or personal data are involved. Classical means for access control lack granularity
when \emph{parts} of the Linked Data graph must be protected. The WebID, combining semantic web concepts with methods from certificate based authentication
and authorization, seems promising to fulfill all requirements concerning security and trust in the semantic web.

In the context of the PerSemID project, we challenged the WebID technology in a \emph{practical scenario} coming from the domain of lifelong learning and 
student mobility. In our use case of study enrollment, we use WebIDs for authentication and to grant access to parts of the triple stores of the
different stakeholders. 
Cross domain triple store interactions are used to exchange data between the involved parties.
Our fully implemented PoC exemplifies an application built on Linked Data and WebID and allows us to judge the usability and security of WebID
technology in a real world scenario.
\end{abstract}




\section{Introduction}
The PerSemID project\cite{persemid} at Bern University of Applied Sciences (BFH) has been set up to
drive the research conducted in the CV3.0 project\cite{cv3} further and to investigate two particular
aspects remaining open in practical applications of the WebID\cite{webid} technology. 

Having a strong focus on security and trust, we have been looking for a validation of WebID specifically 
regarding authentication and authorization. While not questioning its general security properties -- as they are implied by the underlying
mechanisms based on client certificate authentication given by TLS\footnote{Transport Layer Security} -- we have considered
the question of trust, or more specifically the question of \emph{level of assurance (LoA)}\cite{nist,iso29115} in WebIDs to be in need of a more thorough inspection.
The LoA, a concept increasingly used in identity and access management,  states a quality level regarding the authentication -- by which means did it take place and how qualified is it?
As everyone can operate infrastructure to issue WebIDs, the significant question is, what differentiates one WebID from 
another and why (or how) a trust relation to a WebID can be built.
To answer this question, we looked at ways to integrate and/or interlink WebID with other, established systems for
trust and authentication. In particular, we have explored the integration of WebID into SuisseID\cite{suisseid}, Switzerland's 
national electronic ID.

The second open aspect concerns the application of WebID for access control to resources, operated by independent parties
and in distributed environments. Here, we focused on triple stores and platforms for document management.

To conduct the research needed to answer above points as close as possible to real world applications, we have devised
a use case in the domain of lifelong learning and student mobility: enabling Linked Data usage for administrative processes
in enrollment for studies. Here, our work led to a complete implementation as a proof-of-concept prototype (PoC) which can
be used to interactively explore the scenarios which were the basis of our open questions regarding WebID.
Even though being specifically tailored to academic institutions, the concepts developed
in this prototype can easily be adapted to similar processes and needs in other environments.

In this paper, we present the results we have obtained for the given questions. 
After discussing the related work in Section~\ref{sec:relwork}, we describe the basic principles of WebID and detail our work made in strengthening
the LoA in WebID in Section~\ref{sec:webid}. 
In Section~\ref{sec:usecase}, we describe the implemented prototype for distributed resources protected using WebID.
Finally, we conclude our work by summarizing key points and looking forward to future expectations in this area.

\section{Related Work}
\label{sec:relwork}
The PerSemID project lies at the intersection of two domains: identity and access management (IAM), as well as 
Linked Data and semantic web based technologies with a focus on attribute transfer and document management.

Identity and access management in itself is a very broad area covering many aspects in information technology
related to identity, trust, authentication and access. As this project was directly succeeding the CV3.0 project, 
further work based on WebID was directly mandated by the project proposal; due to this, we did not inspect other 
technologies for access management.

We compared our aims to other projects and resources which are closely related to our main objective: having WebID
authentication in cross-domain triple store interaction for Linked Data and documents. 
We will now present the most important related work that we have found.

First, a part of the thesis by Santomauro\cite{santomauro} consists of the SuisseID integration performed in our project. However, the thesis itself is not part of it.

Web Access Control\cite{wac} is one of the first approaches in providing authorization based on WebIDs. Its working principle is based on
access control lists (ACLs) which map WebIDs to HTTP resources and define possible access types like read and write.

Universal Access Control (UAC)\cite{uac}, which we have used in prototypes for CV3.0, goes further and provides access control at the level of individual triples. 

Like UAC, the {Privacy Preference Ontology}\cite{ppo} provides access control at triple level as well.

WebID+ACO\cite{aco} is an ontology for authorization which primarily focuses on adding a role-based authorization model to HTTP.

S4AC\cite{s4ac} is a vocabulary for creating access control policies focusing on named graphs. S4AC is used by the SHI3LD project\cite{shi3ld}
for specifying permissions. SHI3LD is an attribute-based authorization layer with focus on the Linked Data platform and mobile devices.

The former MyProfile project offered an IDP-service for WebID as well as a platform for social networking. At the time of writing,
online resources of MyProfile are not available anymore. However, detailed information can be found in the thesis of Sambra\cite{sambra}.

Recently, a new initiative called Solid\cite{solid} seems to take up on the work of MyProfile. It is also built around WebID as central means of authentication
within a social, Linked Data environment.

\section{WebID}
\label{sec:webid}
In this section, we will first give a short introduction to WebID. We then describe our motivation regarding linking and strengthening WebID and the research
conducted for this. Finally, we will close with an overview of concerns regarding practical applicability of WebID.

\subsection{Basic Working Principle and Motivation}
\begin{figure}[ht]
\centering
\includegraphics[width=8cm]{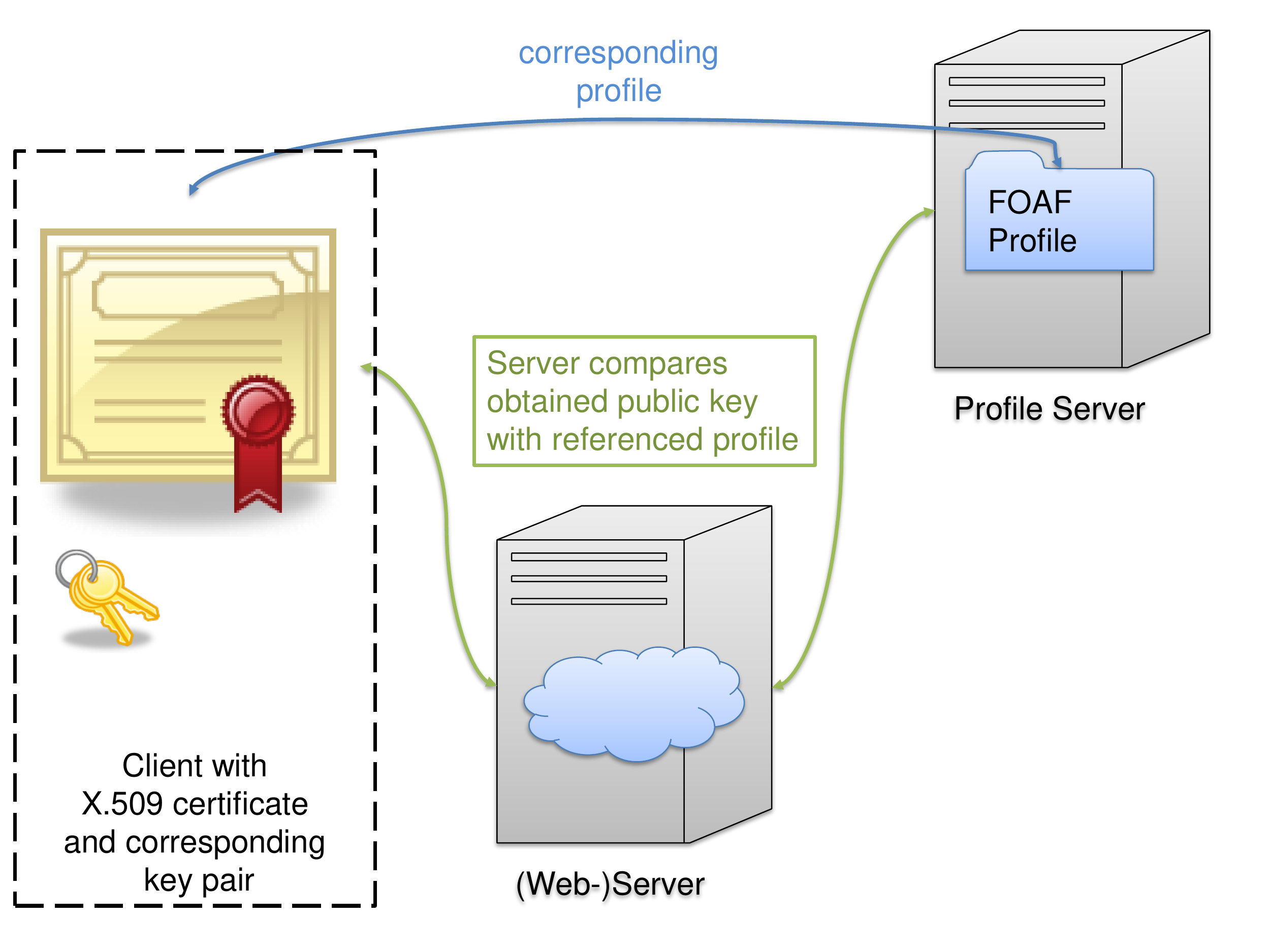}
\caption{WebID working principle}
\label{fig:webid}
\end{figure}

In a nutshell, WebID authentication builds on the authentication of a client using X.509\cite{x509} client certificates. 
Functionality for using such certificates is present in all major browsers (albeit with quite a rudimentary user experience).
To deliver additional information (for example in the form of personal attributes) and to establish a URI as an identificator for a 
particular entity, WebID references a so-called FOAF profile\cite{foaf} using a standard extension of X.509 certificates 
(the \emph{Subject Alternate Name (SAN)} field). Figure~\ref{fig:webid} gives a highlevel overview of authentication using WebID.
On the left, the client (identified by its X.509 client certificate with corresponding key pair) wants to authenticate to
an application running on the (web-)server. The webserver retrieves the FOAF profile referenced in the SAN-field
of the certificate and compares the information about the public key given in it against the information obtained from the 
client certificate in the TLS handshake. If they match, authentication is successful. If desired, additional and potentially signed attributes and other
information can be retrieved from the profile.

As can easily be seen, anyone can issue a self-declared WebID by simply generating an appropriate certificate and publishing a corresponding
FOAF profile document for it on a webserver. \emph{Why should such a WebID be trusted}? Especially in domains which need higher levels of assurance, 
for authentication? Here, clearly the answer is: trust in such a self-declared WebID is not possible without further measures.

The search for trust in WebID is a general problem which can also be found in other systems that use public key
cryptography -- as for instance classical X.509 certificates for websites or secure mail. This problem is known for a long time
and in general, two approaches can be distinguished: the classical, hierarchical PKI\footnote{Public Key Infrastructure}
approach with certificate authorities and the web-of-trust model with peers mutually signing their keys to establish a graph reflecting trust in keys.
However, web-of-trust is more or less constricted to applications of PGP\cite{pgp}, amongst other reasons because X.509 certificates do not allow multiple signatures in
general and there is currently no support for such a mode in all standard implementations. This makes the web-of-trust model thus not suitable for our case.

Additionally, a third, less generally applicable approach would be the linking of identities: coupling an identity with a lower level of trust to another which is more trusted.

We have investigated both ways: linking and adding trust to WebID itself. For adding trust directly, we searched for possibilities to integrate WebID
with established systems having known properties. As trust in X.509 certificates is typically created using the PKI
approach, we have opted for this as well. Detailed descriptions of both approaches will now be given in the following.

\subsection{Interlinking}
\label{sec:interlinking}
\begin{figure}[ht]
\centering
\includegraphics[width=8cm]{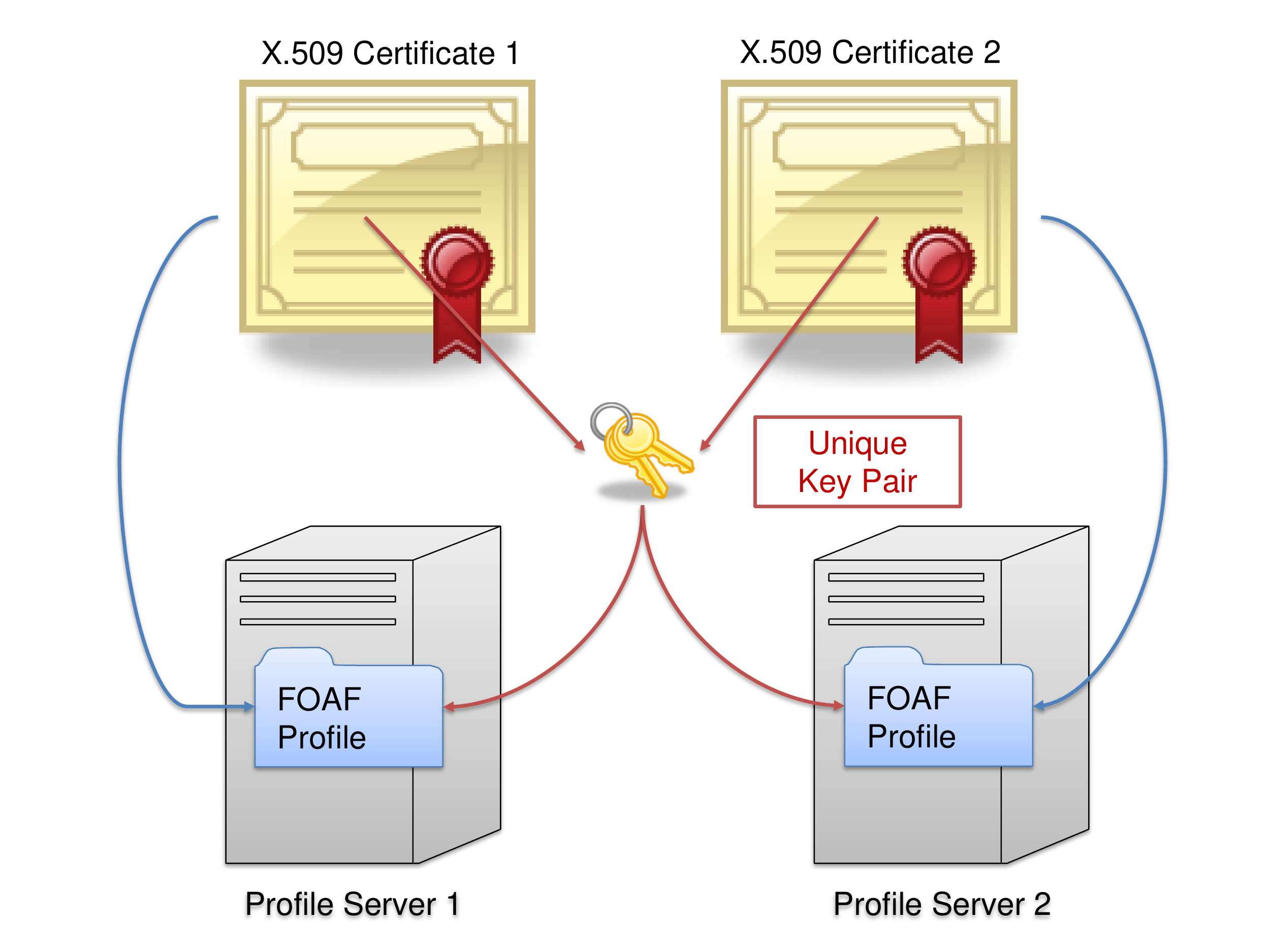}
\caption{Method 1: Linking with unique key pair}
\label{fig:linking-1}
\end{figure}

\begin{figure}[ht]
\centering
\includegraphics[width=8cm]{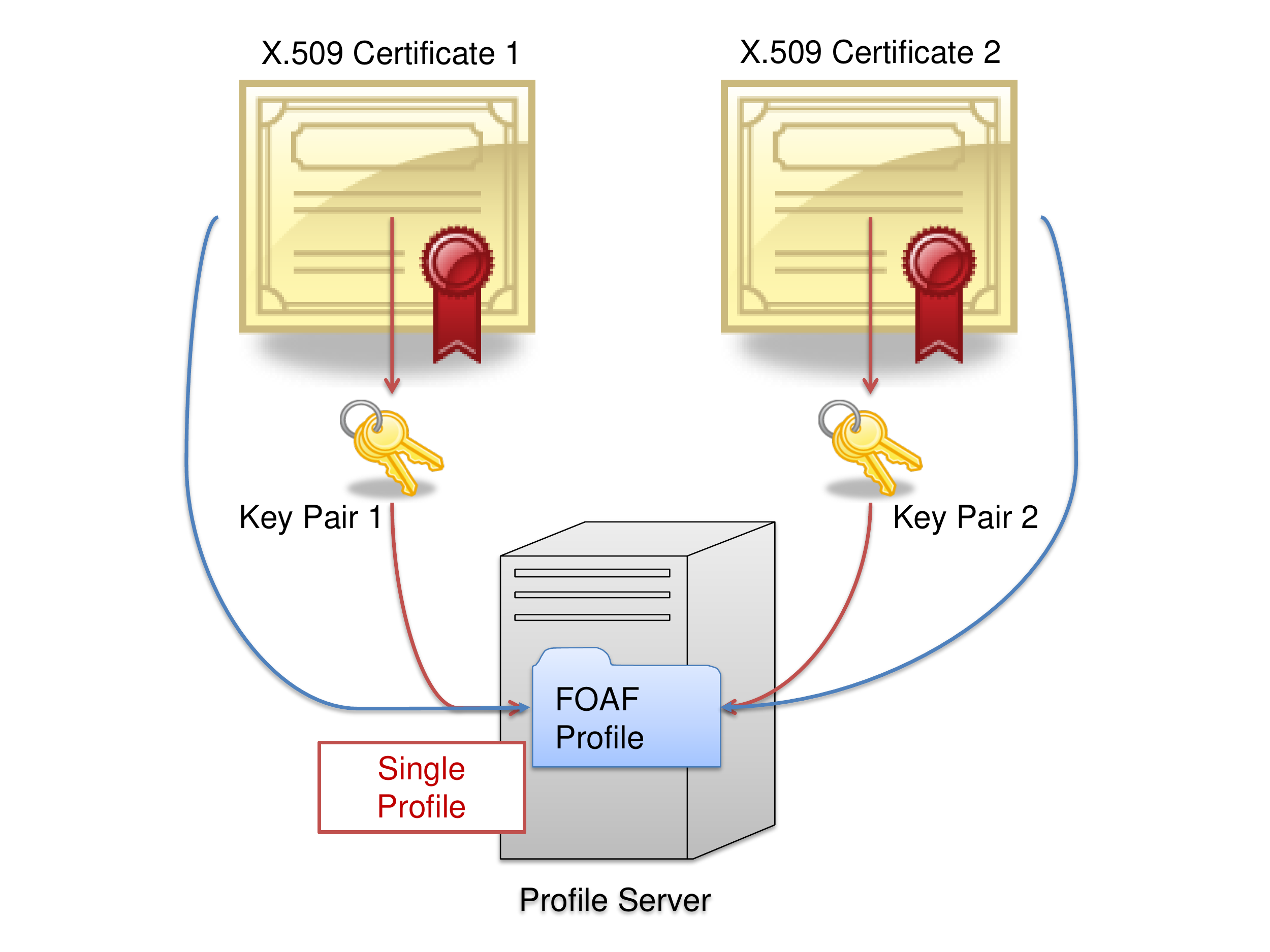}
\caption{Method 2: Linking with single profile}
\label{fig:linking-2}
\end{figure}

\begin{figure}[ht]
\centering
\includegraphics[width=8cm]{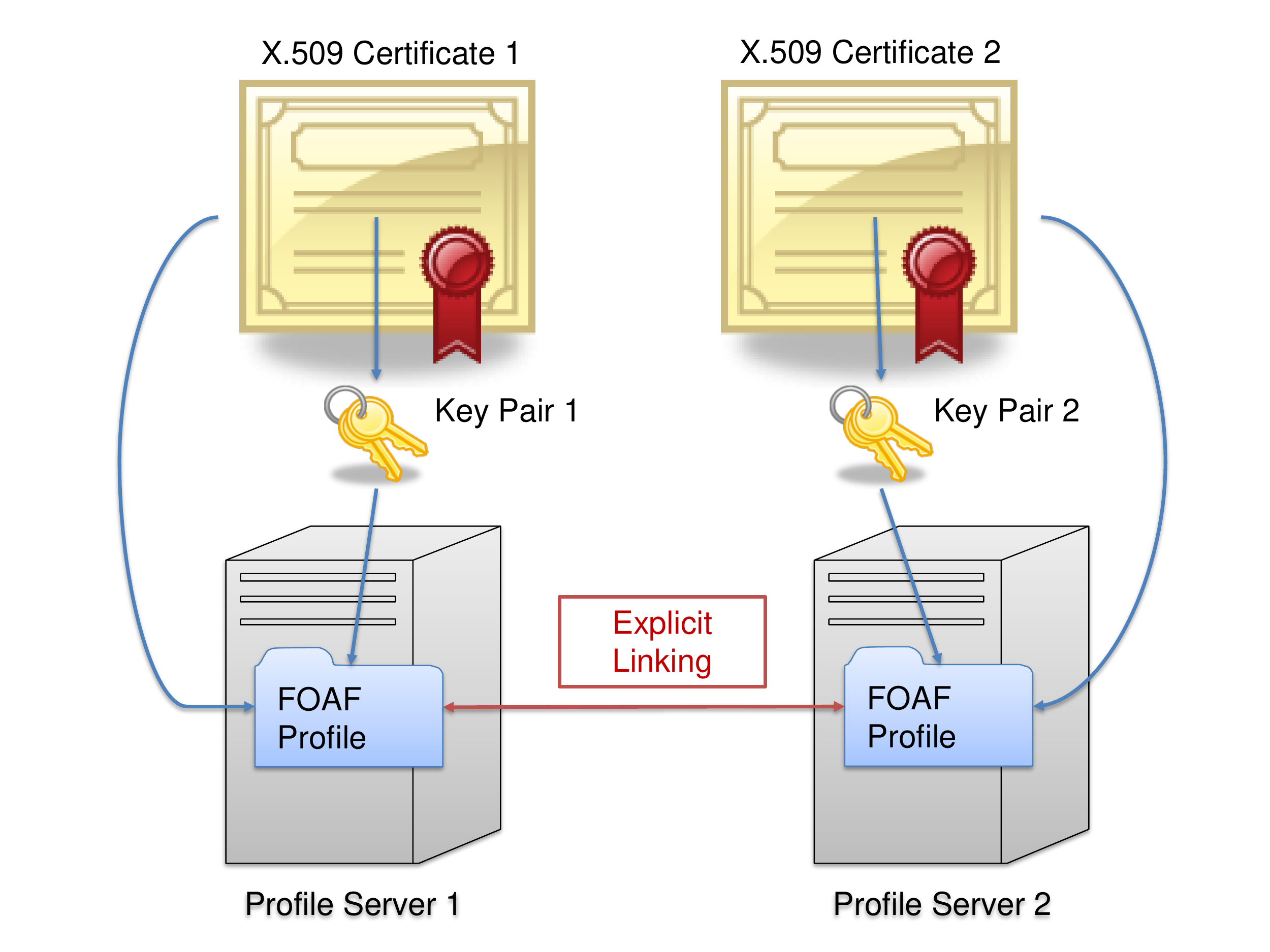}
\caption{Method 3: Semantical linking}
\label{fig:linking-3}
\end{figure}

Interlinking details the process of connecting two identities and declaring them to be about the same subject. In the WebID and PKI context, 
we have devised the following three methods for interlinking WebIDs:

\begin{enumerate}
  \item \textbf{Using a unique key pair shared among certificates (Figure~\ref{fig:linking-1}):} This method uses only a single key pair for which
        two (or more) certificates have been generated. A certificate contains information about the corresponding public key which can be duplicated by a different
        certificate without any issues.
  \item \textbf{With a single profile referenced from multiple certificates (Figure~\ref{fig:linking-2}):} In this case, multiple certificates with
        independent and corresponding key pairs relate to the same FOAF profile URI in their SAN field.
  \item \textbf{By semantically linking key pairs and certificates from a single FOAF profile (Figure~\ref{fig:linking-3}):} Here, again we have multiple
        certificates with independent key pairs. This time, however, each of them points to its own FOAF profile and the relation between certificates
        and profiles is created using additional metadata in the FOAF profile itself.
\end{enumerate}

Depending on the field of application and the restrictions given by the PKI in place, one method may have advantages over
another or may not be available at all. For instance, being given a certificate on a hardware token as with SuisseID, 
sharing a key pair as in Method 1 is not an option. 

The semantical linking in Method 3 is the most flexible but comes with the price that the semantics of the link have to be defined and 
agreed upon initially. In the other methods, semantics are given by the extensions of X.509 certificates and thus are far spread
and well known, making integration and interoperation straightforward.

We think the most practical method for linking identites is Method 2 -- which in turn, however, leaves open the question
of trust as at least one of the involved certificates will probably originate from an untrusted source. We describe a solution
for this issue in Section~\ref{sec:suisseidint}.

\subsection{SuisseID Integration}
\label{sec:suisseidint}
SuisseID, a PKI operated according to national signature laws\cite{zertes} by privately held certificate authorities 
(accredited by the state), provides X.509 certificates for authentication and digital signing on a hardware token.
Besides the certificate, SuisseID also runs an attribute authority which can provide additional information
about the holder of a certificate, like name, date of birth or gender. 

Being widely recognised, accepted, and having a very high level of trust, SuisseID seemed to be an ideal partner for strengthening 
the LoA of a WebID. Furthermore, the attribute authority functionality could seamlessly be integrated into the FOAF profile server,
thus providing the same attributes with the same level of assurance for the Linked Data world.

There are two possible ways to achieve the desired strengthening of WebID using SuisseID: 

\begin{itemize}
  \item Interlinking identities, as described by us in Section~\ref{sec:interlinking}.
  \item Extending SuisseID by adding the needed functionality for WebID to it.
\end{itemize}

Looking at interlinking, we can derive the following: 

Method~1 is not an option for the SuisseID-case, as already stated in Section~\ref{sec:interlinking}, due to the fact that the key lies on a hardware token.
Method~2 is also not applicable, as it would require SuisseID to provide a FOAF profile giving assurance for a key pair generated externally without any control.
As a last option, Method~3 could be applied -- requiring, however, to establish a vocabulary for interlinking identities in a trusted manner between all possible stakeholders, thus reaching too far for our main objectives.

We have thus chosen the option to extend the existing SuisseID with WebID functionality -- we think that this is a very interesting option from different points of view:
Not only would it provide a WebID with the highest level of assurance to each owner of a SuisseID, but it would also help WebID adoption in general,
opening possibilities for a variety of new scenarios in identity and access management.

Also, extending SuisseID with WebID is technically not a hard problem: Certificates issued by SuisseID must be extended to include
the proper subject alternate name extension containing the URI to the corresponding FOAF profile and the issuing certificate authority must operate a webserver for serving these FOAF profile
documents accordingly. 

We have taken on the integration approach as described and validated it prototypically using the demo SuisseID identity provider
which is included in SuisseID's {SDK}\cite{suisseidsdk}.

\subsection{Concerns}
Even though not being a new technology (surfaced end of 2008\cite{msnws}), WebID has not found broad adoption
so far. Having deeply investigated possible applications and issues, we fear that this will not change anytime soon,
unfortunately. \emph{What are the reasons for this?}

WebID, while looking simple and flexible at first sight, suffers from some issues which have been noted by 
others\cite{story,sporny,foafprotocols} before as well. Most notably, the overall user experience of WebID 
seriously hinders broad adoption of the technology. 

This issue not only affects WebID but applications of client  certificates in general. User interfaces for interacting with
certificates in modern web browsers are generally neglected and reduced to the bare minimum; leading to a confusing user experience.
There has been no visible progress or intention on the part of browser vendors to change this in the past and we think, that with
the rise of emerging technologies like {FIDO}\cite{fido}, this fact will not change anytime soon.

Additionally, the world has changed in the meantime and things have become more and more mobile -- or nowadays even things-centered.
Handling of client certificates with multiple browsers on the same operating system has already been cumbersome and tends to get 
impossible with the multitude of devices operated by a single person today; not speaking of ways to perform certificate management on mobile platforms.

\section{Use Case}
\label{sec:usecase}
\subsection{Foundations}
Being focused on applied research, the project team started to investigate possible practical use cases in the second
half of the project. Main goal for this has been to have a live and working implementation of a near-life workflow
in the form of a PoC and thus to be able to challenge the application of WebID regarding our established
requirements. We found the workflows conducted in the enrollment for studies to be an interesting area and have choosen
one specific workflow regarding the enrollment for master studies as our exemplary use case. This workflow involves
three primary actors: A \emph{student} who has successfully obtained a bachelor's degree (and may have additional qualifications), 
the institution at which this degree has been obtained (called \emph{bachelor university}) and finally the institution at which the
student whishes to enroll for master studies (the \emph{master university}).

Business requirements of this use case have been thoroughly analyzed by the project team, leading to a
documentation of the current process in place at BFH's department of business, health and social work
for verifying applications of students for its master program (see \cite{poc}).
This analysis also summarizes attributes and related information concerning metadata about a student's career so far.
The results of it served as the basis for the development of the PoC -- based on it, we have devised application
use cases for all actors leading to a concept for implementation. 

Finally, a fully working prototype has been implemented according to these specifications; the full code\cite{sourcecode}
has been released under the MIT opensource license. We have also produced a screencast\cite{screencast} demonstrating the main 
workflow between all involved parties 

From a technical perspective, PerSemID is a successor of the CV3.0 project and builds upon the concepts of
a personal, semantic curriculum vitae coming out of it. The architecture for a corresponding platform
for serving and maintaining such a CV has been defined in CV3.0's Content Access Service\cite{hitech,cv3cas,cv3bath}.
The \emph{Content Access Service (CAS)} is a RDF triple store with additional document management capabilities as well as
an access control layer.

\subsection{Actors and Their Actions}
\label{sec:actors}
\begin{figure}
\centering
\includegraphics[width=8cm]{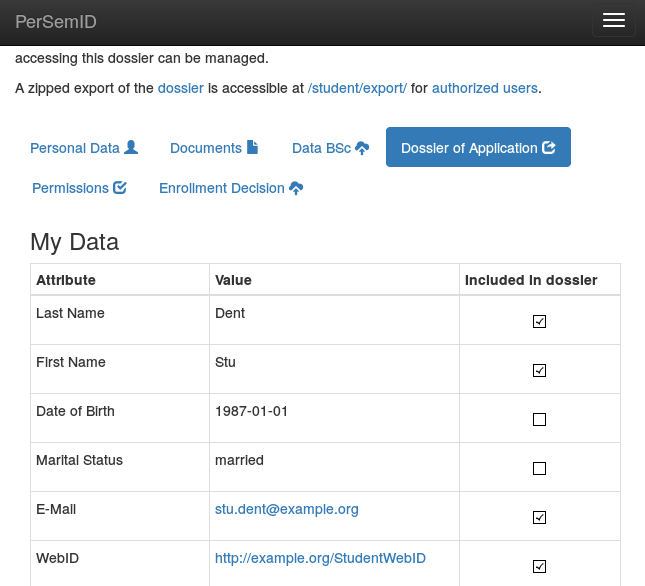}
\caption{Configuration of dossier of application by the student}
\label{fig:dossier}
\end{figure}

In a first step, the student prepares a so-called \emph{dossier of application} which contains all relevant information
about the degree obtained as well as possible additional data in form of documents. Provenance of this data is either personal
information entered by the student directly or data obtained from the bachelor university in the \emph{bachelor dossier}.
The bachelor dossier is issued by the bachelor university as a single file, containing Linked Data about the degree obtained 
and possibly also additional documents. Access to this dossier is limited to the student and it can be interactively downloaded
from the bachelor university. All this data is then stored in the student's CAS and the student can freely 
choose to include/exclude data per application at a master university. A detail of this process can be seen in Figure~\ref{fig:dossier}.

After having created the dossier of application, the student authorizes the master university to access the dossier 
by restricting access based on the university's WebID which is assumed to be publicly available.

Next, the addressed master university picks up the dossier by accessing it on the student's CAS, after having authenticated 
using its WebID. Following a review of all the material in the dossier, a decision regarding acceptance to master studies can be made.
Now, the master university in turn stores its decision on its CAS and authorizes the WebID, given by the student, to access it.

\begin{figure}
\centering
\includegraphics[width=8cm]{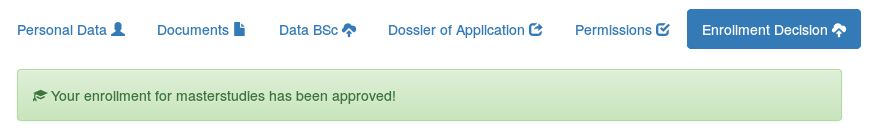}
\caption{Display of enrollment decision}
\label{fig:decision}
\end{figure}

In the last step, the student finally retrieves the decision from the master university, again authenticated by its WebID, and displays it. 
This can be seen in Figure~\ref{fig:decision}.

\subsection{Architecture and Implementation}
Given the three main roles, three distinct applications are needed, one for each of the actors (student, bachelor university and
master university). For ease of implementation and consistency in presentation, they have currently been implemented running
on a single system. Separation is performed depending on the authentication of each actor using a unique, personal WebID.

Being implemented as a classical single page web application, the software consists of a server and a client part. 

\subsubsection{Content Access Service}
To our knowledge, there is no ready-made product similar to a content access service in our notion  as specified by \cite{cv3cas}, thus the needed 
functionality for the PoC had to be implemented in the PoC itself. There is, however, a large range of (mature) triple stores available --
as can be seen for instance in \cite{triplestores}. For our needs, a triple store must support SPARQL 1.1 update\cite{sparql}; together with other
requirements and also based on experience gained in other projects, we have choosen to use Apache Jena\cite{jena}. 

A deliberately reduced set of document management capabilities has been implemented in the PoC-code itself.

The {CAS} serves as storage for all metadata related to each actor and also as location for all application-specific configuration data, like
file system paths or granted permissions. Each actor has their own \emph{named graph} in the triple store
and we perform a copy-process on import of data between different actors. 
An example for the contents of the student's graph, including a granted permission for the WebID \texttt{hmsc.example.org} can be seen in Listing~\ref{listing:student}.

\begin{listing}
\begin{lstlisting}[emph={permission},emphstyle=\underbar]
@base <http://example.org/Student> .
@prefix rdfs: <http://www.w3.org/2000/01/rdf-schema#> .
@prefix xsd: <http://www.w3.org/2001/XMLSchema#> .
@prefix s: <http://persemid.bfh.ch/vocab/student#> .

<#> a s:Student ;
    s:webid <http://example.org/StudentWebID> ;
    s:name "Dent" ;
    s:vorname "Stu" ;
    s:zivilstand "single" ;
    s:geburtsdatum "1990-01-01"^^xsd:date ;
    s:email "stu.dent@example.org" ;
    s:strasse "Examplestreet 3" ;
    s:plz "1111" ;
    s:ort "Exampletown" ;
    s:nationalitaet "Swiss" ;
    s:heimatort "Hometown" ;
    s:wohnortstudba "Studytown" ;
    s:wohnort2jahre "Lasttown" ;
    s:matrikelnummer "1-234-56" ;
    s:sozialversicherungsnummer "123456" ;
    s:permission <http://hmsc.example.org/webid#id> .
\end{lstlisting}
\caption{Example data of a student}
\label{listing:student}
\end{listing}

Data of the other actors looks similar with respective attributes and values. 

Documents, which can be uploaded by the student and the bachelor university, are given a unique ID and stored on the file system.
Metadata needed by the server for interacting with them is again stored in the triple store, in the named graph of the respective actor, an
example is given in Listing~\ref{listing:file}. 

\begin{listing}
\begin{lstlisting}
<http://example.org/Student#7aa5c0f9a76e9a62e3104925c6d6bd81>
  s:fileExtension ".pdf" ;
  s:fileHandle "7aa5c0f9a76e9a62e3104925c6d6bd81" ;
  s:fileName "Curriculum.pdf" ;
  s:fileServerPath "/tmp/psidimas/student/7aa5c0f9a76e9a62e3104925c6d6bd81.pdf" ;
  s:fileSize 605660 ;
  s:fileType "application/pdf" .
\end{lstlisting}
\caption{Metadata of a file}
\label{listing:file}
\end{listing}

Linking to the data of the respective actor is done using the \texttt{file}-predicate of the vocabulary corresponding to the actor.

\subsubsection{Server Application}
The whole server application has been written in JavaScript and is running on node.js\cite{node}. HTTP-functionality has
been realized using the widely deployed middleware layer connect\cite{connect} which makes creation of applications serving
a variety of different requests straightforward. All communication between the frontend application and the server runs 
over a single HTTPS port (HTTPS or more precisely TLS is required for performing the client certificate retrieval needed
for WebID authentication). The following endpoints are served by the server:

\begin{itemize}
\item Webserver for static content (HTML, CSS\dots)
\item Proxy for the triple store
\item Server for the WebID profiles (see Section~\ref{sec:webidp})
\item Document upload
\item Up- and download of ZIP-exports (see Section~\ref{sec:zip})
\item AJAX-actions of frontend
\end{itemize}

Primarily for security purposes (but also for convenience in deployment), connections to the triple store all pass through 
a HTTP-proxy offered as well by the server application.  

\begin{figure}
\centering
\includegraphics[height=6cm]{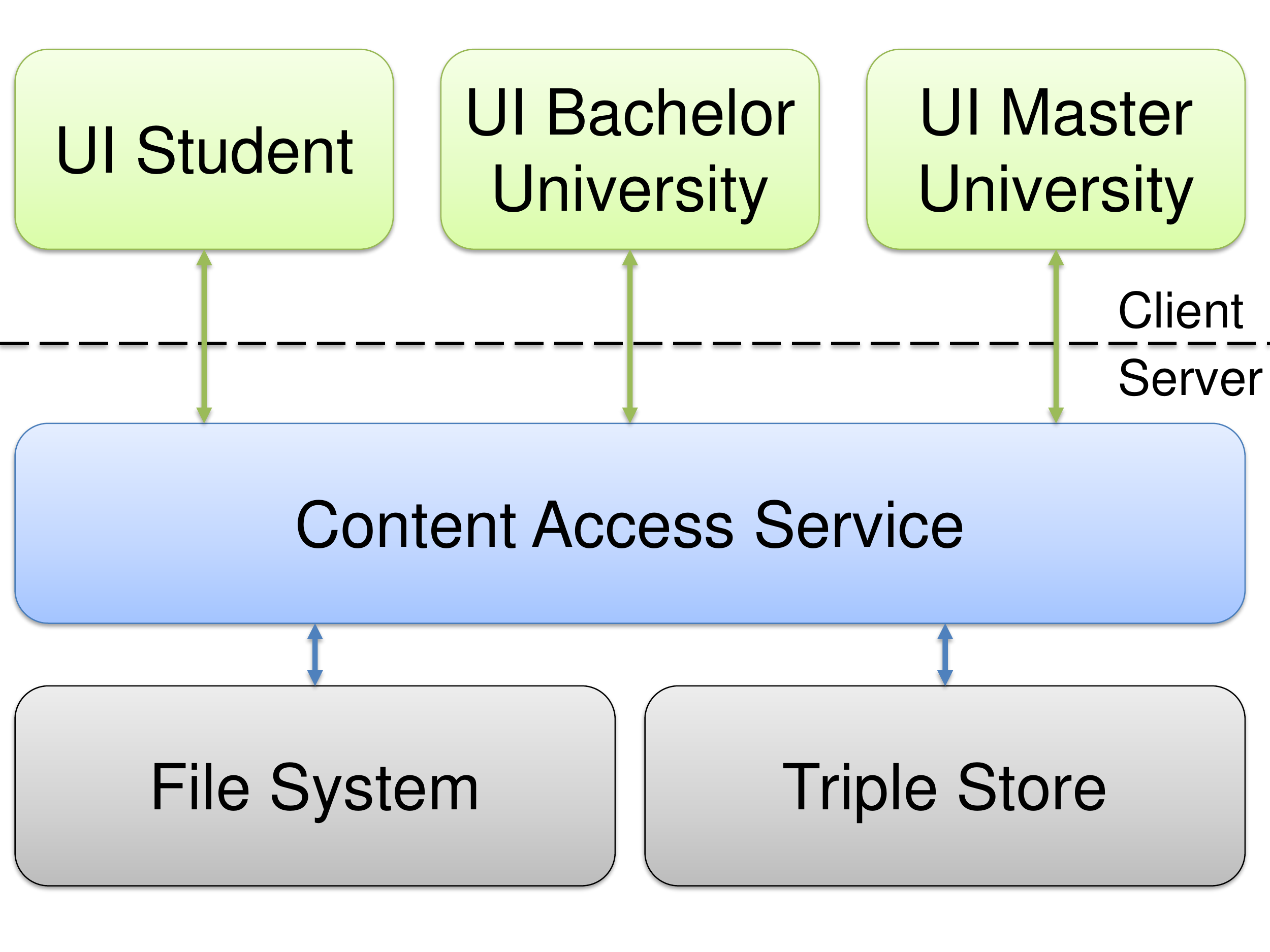}
\caption{Overview of architecture}
\label{fig:architecture}
\end{figure}

On both sides, client and server, we make use of new and advanced JavaScript RDF\cite{rdf} libraries like rdf-ext\cite{rdfext} 
and ld2h\cite{ld2h}. A highlevel overview of the architecture is given in Figure~\ref{fig:architecture}.

\subsubsection{WebID Identity Provider}
\label{sec:webidp}
All functionality needed for WebID authentication has also directly been implemented in the PoC code itself, based on our experience in
the implementation of the WebIDP\cite{webidp} application, an identity provider for WebID developed during the CV3.0 project.
A dedicated URL of the webserver serves the FOAF profiles referenced by the client certificates in use throughout the PoC.
An example of such a profile is given in Listing~\ref{listing:foaf}.

\begin{listing}
\begin{lstlisting}
@prefix cert: <http://www.w3.org/ns/auth/cert#> .
@prefix xsd: <http://www.w3.org/2001/XMLSchema#> .
@prefix foaf: <http://xmlns.com/foaf/0.1/> . 
@prefix rdfs: <http://www.w3.org/1999/02/22-rdf-syntax-ns#> .

<#id> a foaf:Person;
  cert:key [ a cert:RSAPublicKey;
  cert:modulus "c2bcf492 [ ... ] 680f885d"^^xsd:hexBinary;
  cert:exponent 65537 ;
] .
\end{lstlisting}
\caption{Exapmple of a generated FOAF profile}
\label{listing:foaf}
\end{listing}

The certificates for all actors have been generated directly using OpenSSL with respective configuration files.

\subsubsection{Cross-Domain Triple Store Interaction}
\label{sec:zip}

As described in Section~\ref{sec:actors}, all actors follow a defined scheme of interaction. In this scheme, there are three data
exchanges between these actors: download of bachelor data by the student from the bachelor university, download of data from the student by 
the master university and finally, download of data from the master university by the student.

This can be generalized as a concept for sharing data between triplestores or \emph{cross-domain triple store interaction}.
Multiple methods for implementing such interactions could be thought of, we have considered the following three:

\begin{enumerate}
  \item Cross-site sharing using HTTP access control, known as \emph{CORS}\cite{cors}
  \item Proxying of data on the server side
  \item Explicitly channeling data via client-side application
\end{enumerate}

Being limited by the same-origin policy\cite{sop}, that restricts how a document or script loaded from one origin can interact with a resource from another origin, a direct interaction between the client-side program logic and the content access 
service of the remote party in an exchange cannot be implemented -- even considering the fact, that in our PoC scenario, all content was 
served from the same server.  

This problem could be circumvented with HTTP access control (CORS), which allows for a relaxation of the restrictions imposed on the client.
By doing so, we would face another problem: in order to be able to dynamically adjust the needed HTTP headers, parties exchanging data
would have to know each other in advance -- rather an unlikely situation in a real world scenario.

By closely examining this method, one notices that it tends to shift control to the server delivering the client application; so
why not having the server perform the interaction on behalf of the client anyway? Directly acting as a proxy for the data to be exchanged?

Being a seemingly straightforward approach, this method has some serious drawbacks as well. We would have strong concerns regarding
security if the server could be instructed by the client to act as an open proxy interacting with unknown destinations.
Also, for the purpose of our PoC, hiding the exchanges between actors is not optimal for the demonstration of the implemented functionality.

So we finally set with the third option and have implemented a very explicit data exchange using ZIP-files which are downloaded by an actor from
the remote party and manually imported into their own CAS. While this may look odd or even ancient at a first glance, it has some great benefits
for our validation work, which amongst others are:

\begin{itemize}
  \item Explicit WebID authentication and authorization are possible -- our main objective in this case
  \item Separation of the actors and adminstrative borders are clearly visible
  \item Interaction with files is well known to the user
\end{itemize}

We have thus as well implemented the generation of the ZIP file including the necessary contents in our server code.
The basic structure of such a file is as follows: In the root-level directory \texttt{psidimas} (a shortcut used for the PoC),
there are two plain text files: \texttt{data.nt} contains the subset of triples exported from the respective actor's triple store
in Turtle\cite{turtle}-format while \texttt{psidimas.json} is a JSON-file with processing instructions for the import of the data.
Exported documents are stored by using their internal filehandle in the \texttt{files}-subdirectory. An example for this structure is 
given in Listing~\ref{listing:zip}.

\begin{listing}
\vskip 0.15cm
\dirtree{%
.1 export-student|hbsc|hmsc.zip.
.2 psidimas.
.3 psidimas.json.
.3 data.nt.
.3 files.
.4 7aa5c0f9a76e9a62e3104925c6d6bd81.
.4 \dots.
}
\vskip 0.15cm
\caption{Contents of ZIP file exports}
\label{listing:zip}
\end{listing}

\section{Conclusions}
Both parts of the project were able to respond to the specific questions which led to them. The results from these
parts are a significant addition to our previous work on identity and access management intersecting with semantic web and Linked Data
technologies. From our point of view, the project has thus clearly met its objectives.

By investigating means of strengthening the level of assurance in a WebID, we have shown methods for linking of identities with a focus on WebID. This discussion
paved the way for approaching an integration of WebID with SuisseID, showing that the integration only makes sense in the direction of adding
WebID functionality to SuisseID and not the other way round. We have validated these assumptions with an integration of WebID into the SuisseID SDK.

A further, very relevant result and valuable extension of previous work, conducted in the CV3.0 project, was then achieved with the
implementation of the prototype. It has demonstrated, that by using Linked Data technologies, concrete and practical administrative workflows 
can be implemented easily and without hassles. Furthermore, authentication and authorization using WebID stands the test regarding security requirements
in that area -- an integration into other, trusted identification systems such as SuisseID would largely benefit the WebID technology in terms of trust as well.

The prototype also gave us insights in cross-domain triple store interaction and thus provided a model for future implementations of processes and workflows
based on Linked Data technologies. During the implementation, we have also encountered some issues, most notably related to the same origin policy of modern browsers.
For these issues, we have given an overview of possible solutions and described the one chosen by us.

Besides technical problems, our research clearly showed weak points in WebID, some of which have been pointed out by others before.
Regarding broader acceptance of the technology, as means for authentication and especially as \enquote{token} for permission handling, 
future work for better integration, portability and especially userfriendlyness must be undertaken. 
Here, we are particularly interested in approaches taken by recent projects like Solid -- and wether those will be successful in solving these issues.

With the publication of our results, the PerSemID project which ran for 2.5 years comes to an end. Both projects, CV3.0 and PerSemID, fit well in
our institution's portfolio (focusing on identity and access management) and the elaborated results will almost certainly be picked up by future 
research projects.

\bibliographystyle{unsrt}
\bibliography{access-control-in-linked-data-using-webid-full}

\end{document}